\documentclass[12pt]{iopart}
\usepackage{amssymb,latexsym,mathrsfs}
\usepackage{amsfonts}
\usepackage{graphicx}
\usepackage{color}    
\usepackage{textcomp}

\newcommand{\be}{\begin{equation}}
\newcommand{\ee}{\end{equation}}
\newcommand{\bearr}{\begin{array}}
\newcommand{\enarr}{\end{array}}

\newcommand{\ra}{\rangle}
\newcommand{\la}{\langle}

\def\bea{\begin{eqnarray}}
\def\eea{\end{eqnarray}}
\def\ba{\begin{array}}
\def\ea{\end{array}}

\begin{document}

\title{Zero range and finite range processes with asymmetric rate functions}
\author{Amit  Kumar Chatterjee$^1$ and  P. K. Mohanty$^{1,2}$ } 
\address {$^1$CMP 
Division, Saha Institute of Nuclear Physics, HBNI,
1/AF Bidhan Nagar, Kolkata, 700064 India. \\$^2$Max Planck Institute for the Physics of Complex Systems, N \"othnitzer Stra\ss e 38, 01187 Dresden, Germany.
} 
\ead{amit.chatterjee@saha.ac.in; pk.mohanty@saha.ac.in}           


\begin{abstract}
 We introduce and solve exactly  a class  of  interacting particle systems in one dimension where 
particles hop asymmetrically.  In its simplest form,  namely  asymmetric  zero  range process  (AZRP),  particles   hop 
on a  one dimensional  periodic lattice  with  asymmetric  hop rates;  the rates  for both  right and left  moves  depend  only 
on the   occupation  at the departure  site  but their {\it functional forms  are  different.} We  show that  AZRP leads to a factorized 
steady state (FSS) when  its rate-functions  satisfy certain constraints.  We  demonstrate with explicit  examples  that AZRP  exhibits    
certain  interesting  features which were  not possible in usual   zero range process. Firstly,  it can  undergo  a condensation transition  
depending on  how often a particle makes a right move compared to a left one and  secondly, the particle current in AZRP can reverse    its 
direction  as density is changed. We show  that these  features  are  common  in  other  asymmetric models  which have FSS,   like  the 
asymmetric misanthrope process  where  rate functions  for  right and left   hops are different, and  depend  on  occupation of  both the departure 
and the arrival site.  We  also  derive sufficient  conditions for having  cluster-factorized steady states   for  finite 
range process with  such  asymmetric rate  functions  and  discuss possibility of condensation there.
\end{abstract}
\noindent{\bf Keywords: }
Zero-range processes, Non-equilibrium processes, Exact results
\maketitle

\section{Introduction}
\label{intro}
Driven diffusive systems with  stochastic dynamics   have been studied  extensively in recent 
years  to understand  macroscopic  properties  of    non-equilibrium steady states \cite{book1}.
Unlike stationary  equilibrium  systems   which follow   Gibbs  measure,  the  non-equilibrium systems lead to
 unusual steady state measures with interesting  nontrivial  correlations, thermodynamic phases and phase   transitions  even 
in one  dimension   \cite{book2}.   In absence of any generic method for obtaining   exact steady state distributions in 
non-equilibrium systems, the  analytical  
studies  are limited to model systems  where  certain specific    techniques  like,  pairwise balance \cite{pairbalance},   
Bethe  ansatz \cite{bethe},  matrix product ansatz \cite{MPA} etc. can be applied. Zero range process (ZRP), introduced by Spitzer in 
\cite{Spitzer} in context of invariant measures for interacting Markov processes, is one of the   simplest model for which  steady state is 
known    exactly.   In ZRP, particles   hop - one at a time-   to  one of the  nearest neighbors  on a $d$-dimensional lattice   
with a  specific rate function that depends {\it only} on the occupation of the  departure  site,  justifying the name {\it zero range}. 
For   any arbitrary form of the rate function, this model  has a  factorized steady state (FSS) \cite{Spitzer}. In spite of  its   
simplicity, ZRP exhibits  a   condensation transition for certain rate functions,  even in one dimension ($1d$), where a macroscopic fraction of 
particles occupy  a  single site \cite{evans_ZRP}.  In $1d$, such a  condensation transition   can be   mapped   to  a
phase separation in one dimensional exclusion model with suitable  diffusion  dynamics. This   mapping helps in identifying a generic 
criterion for having phase separation in  one dimensional  exclusion models \cite{kafri_phase_sep}. The ZRP  correspondence  of exclusion  models   has 
been exploited  to   obtain   spatial correlation   functions in  several systems \cite{ZRP_exclusion}.

A   natural extension of ZRP  is the finite range process (FRP), where the hop rate of the particles depends  on  the occupation number  of  
not only the departure site but also that of all  other sites within  a  specified  distance \cite{FRP}; clearly in FRP, particle-particle interaction  
extends to  a finite number of neighboring lattice sites.  A specific  example, where the hop  rate   depends on   occupation   numbers  of  both 
the departure and the arrival  sites,  is    commonly known as misanthrope process  (MAP) \cite{evans_beyond_zrp}. Like ZRP, 
misanthrope process  can also have factorized steady state \cite{evans_beyond_zrp},  but only for a certain class of hop rates.    
However, factorized steady state  is not possible  for  FRP \cite{FRP} where  three or more sites are involved in the hop rates. 
For these  systems, in fact,   one  can obtain a cluster factorized steady state (CFSS)   when the  rates  satisfy  certain specific 
conditions  \cite{FRP}. The  simplest  example of a  cluster factorized steady state  is a pair factorized state,  introduced  by Evans {\it et. 
el.}  \cite{evans_hanney_majumdar}, exhibiting  condensation  transition    when   particle  interaction  is tuned.  Interestingly, unlike   systems 
with factorized steady  states    (leading    to a  single site condensate),  in  FRP  condensate can  form over a extended region in the space 
\cite{shape-condensate}  due to  spatial correlations.  Another  interesting variation  is  ZRP   with open boundaries where, in addition  to  the  
ZRP  dynamics  in the bulk, particles  are allowed  to  enter or  exit the system  at the boundaries \cite{open_boundary_zrp}.  
These   open systems   may  not    have   well-defined  stationary states  for any arbitrary boundary dynamics,  but  condensation   can  occur 
for certain   dynamics  which   lead to  unique stationary  measures  \cite{open_boundary_zrp}. Recently, a  non-markovian  zero range  process  
\cite{non_markov_zrp} is  introduced to investigate the impact of temporal correlations on  the  dynamics of condensation.

Over years,   zero range processes have   found vast applications in different areas of science. It  is  being 
considered as a reasonably good model for mass transport processes \cite{mass_transport} and sandpile dynamics 
\cite{carlson, jain},  reconstituting polymers \cite{reconstPolymer} etc.   Being an analytically tractable driven diffusive system, 
ZRP  and related models   have  become  a test ground for  development of  non-equilibrium    thermodynamics \cite{noneq_thermo}.    
These models also help in   understanding  experiments on shaken granular gases \cite{shaken_gg_1}, 
dynamics  of  growing networks \cite{network_1},   aggregation of  active filament bundles  \cite{Kruse}, 
wealth condensation \cite{wealth_con}, jamming in traffic flow \cite{jamming}, quantum gravity \cite{quantum_con} etc. 
Due to their far reaching importance,  ZRP  and related models  have  found 
a  significant place  in  the  research  activity  in  statistical mechanics (see Refs. \cite{evans_ZRP, evans_beyond_zrp} for reviews).

 In usual   ZRP  and related models,   the hop rates  do not depend on the   direction along which  the particles move.  
 Although, recently  some  simple examples \cite{AZRP2d} have been  studied in  two dimension ($2d$), where the  rate functions are  different    in $x$- 
and $y$- directions, but it was observed that the   two point  correlations   are  finite indicating that the steady state  is  not factorized. 
Later,  a  generalized zero range processes  was introduced  \cite{Lebowitz}  where   more than  one particle  can  hop  from a   site 
 and the hop rates   may depend on direction of hopping. 
 A sufficient condition for having    FSS in these models, which is also conjectured  as  the  necessary  condition,   showed explicitly 
 that  indeed   models  described in \cite{AZRP2d} cannot have  factorized steady states. 
Moreover, these  models in   $1d$ (with  one hop at a time)  reduce   to  an asymmetric ZRP where  particles hop to  right 
 or left neighbour with rates  $u_R(n)= p u(n), u_L(n)=  q u(n)$ respectively;  notably, the steady state weights  of these  models   
do not depend  on $p, q$ and the  asymmetry  parameter $\frac{p}{q}$  only redefines the fugacity  of the system in grand canonical ensemble.

In this article  we introduce  a class  of   one dimensional  interacting   particle  systems  with {\it asymmetric rate functions}, 
i.e.,  the right hop rate   $u_R(n)$  is an   independent function, not just a  constant multiple 
of  the left hop  rate $u_L(n).$   It is {\it a priori}  not clear,     whether   a  factorized steady  state  is 
at all possible for  this  asymmetric zero range process (AZRP).  We   derive   a  sufficient   condition   for  AZRP  to have 
a   factorized steady state.   Generalization of these  asymmetric   models   to  asymmetric misanthrope process (AMAP) and asymmetric 
finite range process (AFRP)  are also  investigated to  find  sufficient conditions on the rate functions   that lead  to  factorized steady state 
in  AMAP and  cluster factorized form  for  AFRP. Interestingly, even though the  steady state of  both  AZRP and  AMAP are similar 
to that of ZRP,    particle currents   here  show  current-reversal  as  the  density of the system  is changed  - a feature 
which can not be   observed  in  ZRP with rates  $u_R(n)= p u(n), u_L(n)=  q u(n).$   
We also address the possibility  of condensation transition in  these  systems and  
find that the onset of condensation can be tuned by the a factor  that  merely controls  
how often the  particle  chooses to move    right,     compared to its left hops.

The asymmetric  hopping models which  we   discuss in this article are interesting in their own right.  In addition, 
there are  physical situations  which may correspond to the asymmetric  diffusion  proposed  here.  It is well  known that  
geometry  \cite{Geometry} or  potential of mean forces \cite{BioChannel} induce  asymmetry  across membrane channels and influence the particle fluxes  
across artificial or natural-biological pores. Such  asymmetry   is important for analyzing the dynamics of particle translocation 
\cite{Kolomeisky} in biological channels. Also, this asymmetric diffusion effect may  be utilized \cite{Bacteria} to regulate 
transport and distribution of motile microorganisms in irregular confined environments, such as wet soil or biological tissues.

The article is organized as follows. In  section  \ref{sec:1}, we introduce  AZRP  and derive the  sufficient condition 
on the rate functions for obtaining  a FSS. We   then calculate the generic form of these  rate functions  $u_R(n), u_L(n)$  that  gives rise to FSS, we 
 devote  the  rest of the section   for elaborate discussions on phenomena of condensation and current reversal. 
 In section  \ref{sec:3}  we  introduce   asymmetric misanthrope process  and show that the system can  lead to 
 FSS   under  certain   conditions;  current reversal and condensation  phenomena in AMAP are discussed with  specific examples. 
 The most generic case, asymmetric   finite range process (AFRP)   is discussed in section   \ref{sec:4}, which    leads  to 
 a cluster factorized steady states as  in  \cite{FRP}.   Finally,  we summarize the results   in section  \ref{sec:5} with   
 some discussions.\\

\section{Asymmetric  zero range  process (AZRP)}
\label{sec:1}
\subsection{\textbf{ The Model}  }
\label{sec:2} Let us consider a system of $N$ particles on a one dimensional periodic lattice with $L$ sites  labeled by 
$i=1,2,\dots,L .$ Each site $i$ can accommodate $n_i \geq 0$ number of particles.  The dynamics of the system  is as follows. 
From a  randomly chosen site $i,$  having  $n_i >0$ particles,  one  particle  is   transferred  either   to the right neighbor $(i+1)$
with a rate   $u_R(n_i)$ or to its left neighbor $(i-1)$ with a different rate function $u_L(n_i)$. Thus,  the total 
number of particles $\sum_{i=1}^{L}n_i=N$ or  the density $\rho= N/L$   is  conserved. 
This stochastic process is a zero range process with asymmetric rate functions and hereafter we refer to it in 
short as asymmetric zero range process (AZRP).  Clearly, in AZRP, particles at any  given lattice 
site  interacts with other particles at the same site through the hop rates  which  explicitly depend on the 
occupation number; interaction between particles at different sites is invoked only via the 
global conservation of $N$.  In the following we show that this  interacting particle system  can  have a factorized steady 
state if the rate functions  satisfy certain constraints.

A special case of the model with $u_R(n_i)=  p u(n),  u_L(n_i)=q u(n)$  is the   well  known zero range  process \cite{evans_ZRP} 
which describe   symmetric  (when $p=q$)   or  asymmetric  (when $p\ne q$) transfer of particles. In this case, the  steady 
state has a  factorized form for  any choice of rate function  $u(n)$, and for arbitrary  values of  $p,q$
  \be
       P_N(\{n_i\})  \sim \prod_{i=1}^{L} f(n_i) \delta(\sum_{i=1}^{L}n_i-N), 
   \label{eq:FSS}
  \ee
where $f(n) = \prod_{m=1}^{n}u(m)^{-1}.$
We now ask, if such a  factorized form   is possible when  rate functions   for  right and left hops  are   different, i.e., $u_R(n)$ 
and $u_L(n)$ have distinct functional forms.   The  master equation  for   AZRP is    
\bea
\frac{d}{dt} P(\{n_i\})=\sum_{i=1}^{L} \left[ u_R(n_i)+u_L(n_i) \right] \,\,P(n_1,\dots,n_{i-1},n_i,n_{i+1},\dots n_L) \cr
 -\sum_{i=1}^{L}\left[u_R(n_{i-1}+1) P(\dots n_{i-1}+1,n_i-1,n_{i+1}\dots )\right. \cr \left.+ u_L(n_{i+1}+1)P(\dots n_{i-1},n_i-1,n_{i+1}+1\dots)\right]
\label{eq:master_AZRP}
 \eea
 which governs   how    the  probability $ P(\{n_i\})$   of  configuration $\{n_i\}$    evolves with time.
Let    us  assume  that the steady state  of AZRP   has a   factorized  form, as  in  Eq. (\ref{eq:FSS})- then we use the 
FSS in Eq. (\ref{eq:master_AZRP})  to check whether the steady state condition  $\frac{d}{dt}P(\{n_i\})=0$ is satisfied 
automatically or does it put some constraint on $u_{R,L}(n)$  for which FSS  is possible.  
With  a FSS,  the steady state master equation for any arbitrary configuration of AZRP reads  as, 
\be
\bearr{c}
\sum_{i=1}^{L}  \left[u_R(n_i)+u_L(n_i)\right] \,\,f(n_1)\dots f(n_{i-1})f(n_i)f(n_{i+1})\dots f(n_L) \, \\
 -[\,\sum_{i=1}^{L} u_R(n_{i-1}+1)\dots f(n_{i-1}+1)f(n_{i}-1)\dots \cr +\sum_{i=1}^{L} u_L(n_{i+1}+1)\dots f(n_i-1)f(n_{i+1}+1)\dots]=0. \\
 \\
\enarr
\ee
Now by shifting the index $i \rightarrow (i-1)$ in the last sum we get an equation 
$\sum_{i=1}^{L} F(n_{i-1},n_i)=0, $ where
\bea
F(m,n)&=&u_R(n)+u_L(n) - u_R(m+1) \frac{f(m+1)f(n-1)}{f(m)f(n)}\cr
&&- u_L(n+1) \frac{f(m-1) f(n+1)}{f(m)f(n)}.
\label{eq:Fmn}
\eea  
Clearly  we have a  stationary measure  if  we  can construct  a single site function $h(n)$ that  satisfy  
$F(m,n) = h(m) - h(n).$   Existence  of such a function $h(n)$   ensures  that  $\sum_{i=1}^{L} F(n_{i-1},n_i)=0$ 
and thereby  guarantees  a  factorized  stationary measure. 
Since $m,n$ are  non-negative  integers,  let  us first find   what    restrictions  are   imposed on $h(.)$  from  
the  boundary values. When $m=0=n,$  from Eq. (\ref{eq:Fmn})  we  have  $F(0,0) =0$, as $u_{R,L}(0)=0$ (particle  hopping is prohibited 
 if   the  departure site  is vacant) and $f(-1)=0$  (a boundary condition that assigns zero  weight  for  configuration   having  $-$ve occupation  
 numbers);   thus  $F(m,n) = h(m) - h(n)$ is  automatically satisfied. For other  cases,     
\bea
  n=0, m>0&:&~ -u_L(1) \frac{f(m-1) f(1)}{f(m)f(0)}=h(m)-h(0) \cr
  n>0, m=0&:&~ u_R(n) + u_L(n)- u_R(1)\frac{f(n-1) f(1)}{f(n)f(0)}=h(0)-h(n).
 \label{eq:condition_1}
\eea
These  equations  are   consistent if 
\be
 \hspace*{-2 cm} f(n) = \frac{f(1)}{f(0)} [\frac{u_R(1)+u_L(1)}{u_R(n)+u_L(n)}] f(n-1), ~{\rm and}~ h(n)=h(0)-u_L(1) \frac{f(n-1) f(1)}{f(n)f(0)}.
 \label{eq:AZRP_FSS0}
\ee
 Finally, a  factorized steady state  will be guaranteed if   the above expressions of $h(n)$ and  $f(n)$  consistently satisfy  $F(m,n) = h(m) - h(n)$ 
for all  $m>0,n>0.$   This  requirement actually constraints 
the right and left hop rates $u_{R,L}(n)$ to  satisfy the following condition (from Eqs.  (\ref{eq:Fmn}) and (\ref{eq:AZRP_FSS0})) ,
\be
 \frac{u_L(n+1) u_R(1) - u_R(n+1) u_L(1)}{[u_R(n) + u_L(n)] \, [u_R(n+1) + u_L(n+1)]} = C, \\
 \label{eq:AZRP_constraint}
\ee 
where $C$ is a constant independent of $n$.
This   completes  the   proof: {\it AZRP   has a factorized   steady state  if the hop rates $u_{R,L}(n)$  satisfy 
Eq. (\ref{eq:AZRP_constraint})}. The  weight factors $f(n)$  can be calculated   from the   recursion  relation 
Eq. (\ref{eq:AZRP_FSS0}) 
 \be
 f(n) = [f(1) v(1)] ^n \prod_{m=1}^n  \frac{1}{v(m)} ~;~{\rm where}  ~ v(m)  = u_R(m)+u_L(m) , 
 \label{eq:AZRP_FSS}
 \ee
 where  we set  $f(0)=1$, without loss  of generality.   Note a  striking similarity   of  the   weight   factor $f(n)$ in AZRP with 
 that of  the ZRP.   In Eq. (\ref{eq:AZRP_FSS}) if one sets $f(1)=\frac{1}{v(1)}$, then the  steady state   of  AZRP   with  specified   hop rates $u_{R,L}(n)$ 
 which  satisfy Eq. (\ref{eq:AZRP_constraint}) is exactly the same  as that of the  ordinary  ZRP with hop rate $ u_R(n)+u_L(n).$
 
 Note that, although   validity of  Eq. (\ref{eq:AZRP_constraint}) is  sufficient  for AZRP to have a  FSS, it is not 
{\it  a priori} clear if  there  exists any such rate functions which satisfy this condition. To obtain a desired FSS as 
 in Eq. (\ref{eq:FSS})  where  
  \be
  f(n) = \prod_{m=1}^{n} \frac{1}{v(m)} \:\:\: \mathrm{along with} \: f(0)=1
  \label{eq:f_v}
 \ee
one can show, following  Eqs. (\ref{eq:AZRP_FSS})  and (\ref{eq:AZRP_constraint}), that  the asymmetric rate 
 functions  have the following generic functional form for $n \ge 1$,
 \bea
 u_R(n) = v(n) \left[ \delta - \gamma  v(n-1)\right]~;~ u_L(n) =   v(n) \left[ 1- \delta + \gamma  v(n-1)\right].
 \label{eq:AZRP_generic_form_rates}
\eea
Clearly for $n=0,$ $u_R(0)=0=u_L(0)$ meaning $v(0)=0$. Also we have set $\frac{C}{v(1)}=\gamma$. Now we   have 
a family of  asymmetric hop rates,  characterized by  two independent  
parameters $0\le \delta \le 1$ and  $0 \le \gamma\le \delta/ v(n)|_{max}$ \footnote{The  range 
of $\delta$ and $\gamma$  are fixed  by the condition that  the rates $u_{R,L}(n)$  must  be   positive.},
which  gives rise    to a  unique  invariant   measure  described  by  Eqs. (\ref{eq:FSS}) and  (\ref{eq:f_v}).

Some specific  examples of AZRP will  be discussed in  the following  sections. A simple   situation  is 
when $\gamma=0,$ where    $u_R(n) = \delta v(n)$  and $u_L(n) = (1-\delta) v(n).$  Since $\delta<1,$    
the model is   identical to an  ordinary ZRP  where  particle  
chooses  the right (or the left) neighbor as a target site   with probability $\delta$ (or $1-\delta$)
and then  hops to that site with rate  $v(n).$ Obviously, $\delta=0,1$ corresponds  to   the usual
 ZRP  where  particles hop along a  unique direction.

For  any conserved system ($N$ particles  in $L$ sites)   with a factorized steady state
\bea
 P_N(\{n_i\})  = \frac{1}{Q^L_N} \prod_{i=1}^{L} f(n_i) \delta(\sum_{i=1}^{L}n_i-N), ~{\rm with}~  f(n) = \prod_{m=1}^{n} \frac{1}{v(m)},  \label{eq:FSS2}\\
 {\rm where} ~ Q^L_N =\sum_{\{n_i\}} \prod_{i=1}^{L} f(n_i) \delta(\sum_{i=1}^{L}n_i-N)
 \eea
is the canonical partition function, one can  calculate  the steady state  average of any local observable  straightforwardly.
For  completeness let us describe  the procedure briefly.  The grand partition function of the system is 
\be
Z_{L}(z) = \sum_{N=0}^\infty Q^L_N z^N = F(z)^L; ~~ F(z) =  \sum_{n=0}^\infty f(n)  z^n , 
\ee
where   the fugacity $z$ controls  the  average  density  of the system $\rho(z) = z F'(z)/F(z).$   The  steady state  average value of any local observable 
$O(n_i)$  is then 
\be \la O \ra=\frac1{F(z)}  \sum_{n=0}^\infty  O(n) f(n) z^n,\ee which is a  function of $z.$  One can get the corresponding  value   for the conserved system 
with a given density  $\rho=\rho^*$  by  setting   $z$  to a  specific   value  $z^*$   which satisfy  $\rho(z^*)=\rho^*.$  

\subsection{\textbf{Condensation}}
The  most interesting   thing that  happens  in ZRP with a   hop rate $v(n),$ or  for any other model which has a 
factorized steady state given by Eq. (\ref{eq:FSS2}), is the condensation transition. If  the asymptotic form of $v(n)$  is  
\be
v(n) =  v(\infty) \left(  1 +   \frac{b}{n^\sigma} + \dots \right),
\label{eq:rate_ZRP_cond}
\ee
condensation occurs  for  large densities  either   when $\sigma<1,$ or when $\sigma=1$ and  $b>2$   \cite{evans_ZRP}.
It   turns  out  that  higher order  terms  in the  series expansion  are  irrelevant  in deciding  the  possibility of a  condensation transition;  
they only  play a role  in determining  the  exact critical density above which  the system  forms a condensate. 
Since    there  are   many exclusion models that have  exact or approximate  ZRP correspondence, 
the  above criteria is extensively  used  for determining  the  possibility  of  phase separation transition 
\cite{kafri_phase_sep}.  A particularly simple case of (\ref{eq:rate_ZRP_cond}), which is exactly solvable 
\cite{evans_ZRP}, is \be  v(n) = 1+\frac{b}{n} \label{eq:1bn} \ee   that results  in a  condensation  
transition   for $b>2$, when  density $\rho$ of the system crosses a critical value  $\rho_c = \frac{1}{b-2}.$

In AZRP, to have  a  FSS   given  by  (\ref{eq:FSS2})  with $v(n) = 1+\frac{b}{n}$ for $ n\geqslant 1$  
($v(0)=0$ by definition as already mentioned) the rate functions   must  follow Eq. (\ref{eq:AZRP_generic_form_rates}). 
For  this choice of $v(n),$ 
the  model  has three parameters $b>0,$ $0< \delta \le1 $  and    $\gamma;$ here   $\gamma$ must   
be in the range $0 \le \gamma  \le    \frac{\delta}{v(n)|_{max}}=  \frac{\delta}{1+b},$ so that  the  rates   in 
Eq.   (\ref{eq:AZRP_generic_form_rates})  remain  positive for all $n>0$. Let us   parametrize  $(b, \delta,  \gamma)$  in  
terms of  three other parameters $(b_R, b_L, \alpha)$  as follows,   
\bea 
b=\alpha b_R + \bar \alpha b_L~;~~  \delta =  \alpha  ( 2- \frac{b_R} {\alpha b_R +\bar \alpha b_L})~; ~~
\gamma =  \alpha  ( 1- \frac{b_R} {\alpha b_R +\bar \alpha b_L}),
\label{eq:dgb}
\eea
where    we  use $\bar \alpha \equiv 1 - \alpha$ for  notational convenience. 
The purpose of such parametrization  will become clear  in a moment.  With these  new parameters   the  
hop  rates of the  model   for  the choice  $v(n) = 1+\frac{b}{n}$   can be written (using    Eq. (\ref{eq:AZRP_generic_form_rates}))  as
 \be 
 u_R(n) = \alpha \tilde u_R(n), u_L(n) = \bar \alpha \tilde u_L(n)
 \label{eq:arl1}
 \ee
where  for $n=1,$ 
\bea
\tilde u_R(1) = (2-\frac{b_R}{\alpha b_R + \bar \alpha b_L})\left[ 1+\alpha b_R + \bar \alpha b_L \right] \cr 
\tilde u_L(1) = (1-\frac{b_R}{\alpha b_R + \bar \alpha b_L})\left[1+\alpha b_R + \bar \alpha b_L \right] 
\label{eq:arl2}
\eea
and for $n>1$
\bea
\tilde u_R(n) =  (1+\frac{\alpha b_R + \bar \alpha b_L}{n})\left[1-\bar \alpha \frac{b_L-b_R}{n-1}  \right] \nonumber\\
\tilde u_L(n) = (1+\frac{\alpha b_R + \bar \alpha b_L}{n})\left[1+\alpha \frac{b_L-b_R}{n-1} \right].
\label{eq:arl3}
\eea
It is easy to  see that  the   asymptotic forms of $\tilde  u_{R,L}(n)$  are 
\be
\tilde u_{R}(n) =1+   \frac{b_{R}}{n} + \dots  ~;~~~\tilde u_{L}(n) =1+   \frac{b_{L}}{n}+ \dots  .
\ee

The new parameters    $\alpha,b_R,b_L$    are  all  familiar  to us:  $b_{R,L},$  are  coefficients of  $\frac{1}{n}$  in the asymptotic
expansion of  the rates   $ \tilde u_{R,L}(.)$   which   normally  take part in  determining   possibility  of a condensation transition,  and 
$\alpha$ may  be considered  as the probability that   a particle   chooses   the right neighbor as the target site
(note that $\alpha= \gamma-  \delta$ varies  in the  range $(0,1)$ for any $b>0$).   Thus, for  the  model in  hand,  
particles  choose    to move right (or left)   with probability $\alpha $ (or  $1-\alpha$)  and   hop there  with rate $ \tilde u_{R,L}(.)$ respectively.

For $\alpha =0,$    particles   in this model  move  only  to left   with rate $\tilde  u_L(n) = 1+ \frac{b_L}{n}$  leading to a 
factorized steady state and  a  condensation  for large densities when  $b_L>2.$ Similarly   for $\alpha=1,$  
condensation occurs for $b_R>2.$  It is interesting to ask,  
$`$for a  given  fixed $b_{R,L},$ is it possible to  observe a  condensation transition  by changing $\alpha$ ?'   Note that $\alpha$   determines 
how often the system chooses  to hop right and a condensation   transition,   if  appears  by  tuning only  $\alpha,$   is   exciting as
it  has not been observed earlier  in  ZRP  or related models.

The difficulty, however, lies  with the fact  that  for any given $b_{R,L}$ we  do not   have  exact  steady state  measure  (within  this  formalism) 
for all $\alpha \in (0,1)$
The constraint  comes from the  requirement  that  the  rate functions  obtained in Eq. (\ref{eq:arl1})-(\ref{eq:arl3})  must be 
positive valued  for  $n>0,$
which in turn restricts the  value  of $\alpha$  for which  one  can obtain  the steady state  weights exactly.  In other words, 
for  some $b_{R,L},$ it may not be possible   to  find  $u_{R,L}(n)$  for which the steady state  is factorized  for 
any arbitrary  $0\le \alpha\le 1.$  
When  both $b_R$ and $b_L$ are   larger than $2,$  we have   $b= \alpha b_R + (1-\alpha) b_L >2;$ 
this case is not interesting because,  even if  we find  suitable hop rates that  describe this situation,  and  result in  a FSS  
as in Eq. (\ref{eq:FSS2}) with $v(n) = 1+ \frac b n,$   the  system will  remain  in the condensate  
phase for all  $\alpha.$  Similarly, for $b_R<2,b_L<2,$  condensation transition is not possible as $b$ is smaller than 
$2$ for any $0<\alpha<1.$   Thus, we focus on the case where $b_R<2$  and $b_L>2$  (the other alternative  $b_R>2$  
and $b_L<2$ can be described in the same manner).  
For  any fixed value of $b_R$  the  minimum   and the maximum accessible values of  $\alpha,$  for which  one can  
have exact FSS  with rate functions $u_{R,L}(n)$ given by Eq. (\ref{eq:arl1})-(\ref{eq:arl3})  are respectively
\be
\alpha_{min}=  \max \{0, \frac {b_L-b_R-1}  {b_L-b_R}\}  ; ~~  \alpha_{max}=  \min \{1, \frac12\frac {b_L}{b_L-b_R}\}.
\label{eq:alpha_range}
\ee
These conditions  on $\alpha$  are  calculated  simply by demanding   positivity of the hop rates in (\ref{eq:arl1})-(\ref{eq:arl3}).

To demonstrate  the possibility of  a   condensation transition tuned by   $\alpha,$  we consider  AZRP with   hop rates 
$u_{R,L}(n)$  
given by  (\ref{eq:arl1})-(\ref{eq:arl3}),   in two separate cases $b_R=\frac32$  and   $\frac12.$    
The  maximum and minimum values of $\alpha$ now depends on $b_L;$
\begin{figure}[h]
\vspace*{.5 cm}
 \centering
\includegraphics[height=5.5 cm]{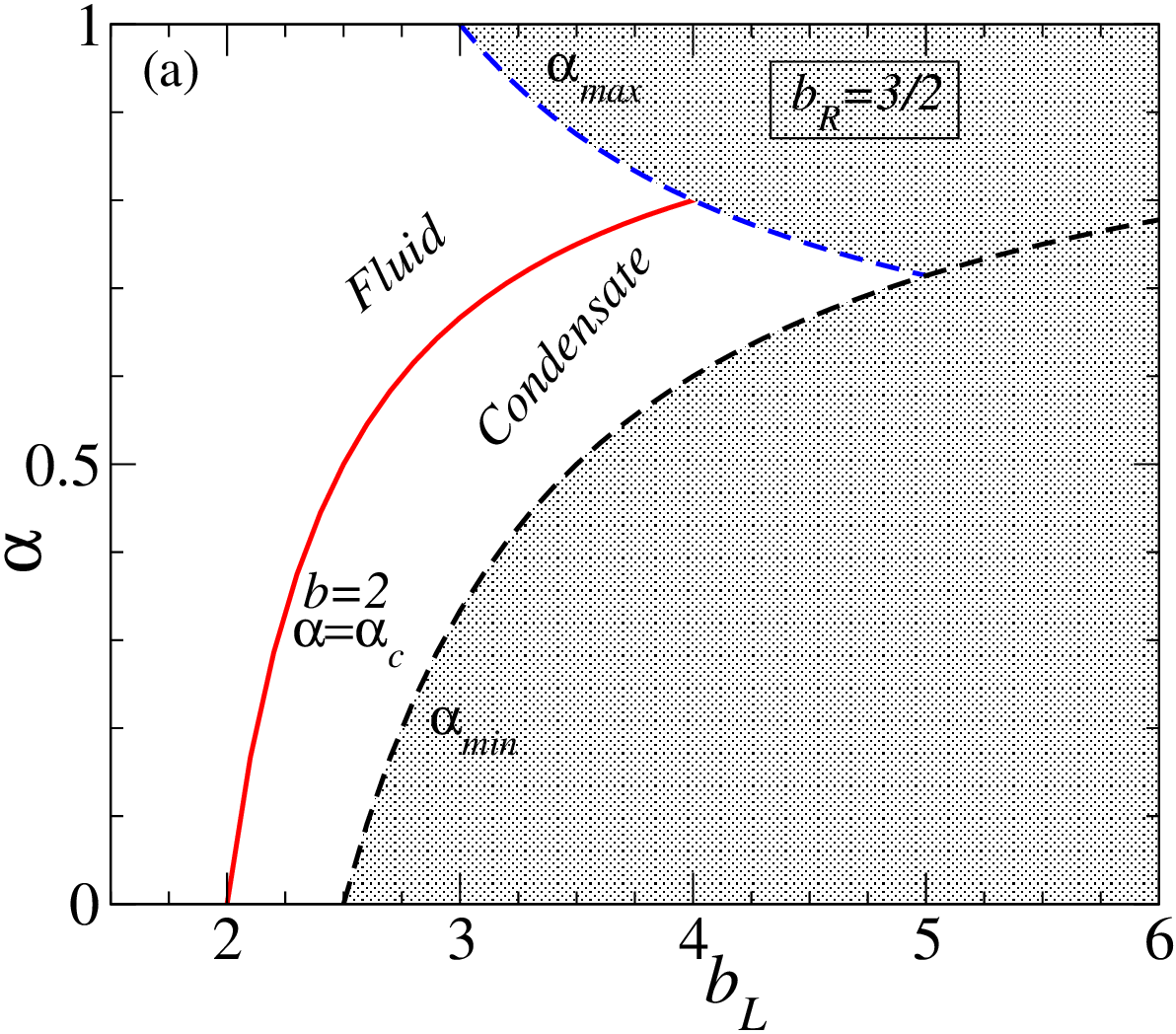}\includegraphics[height=5.5 cm]{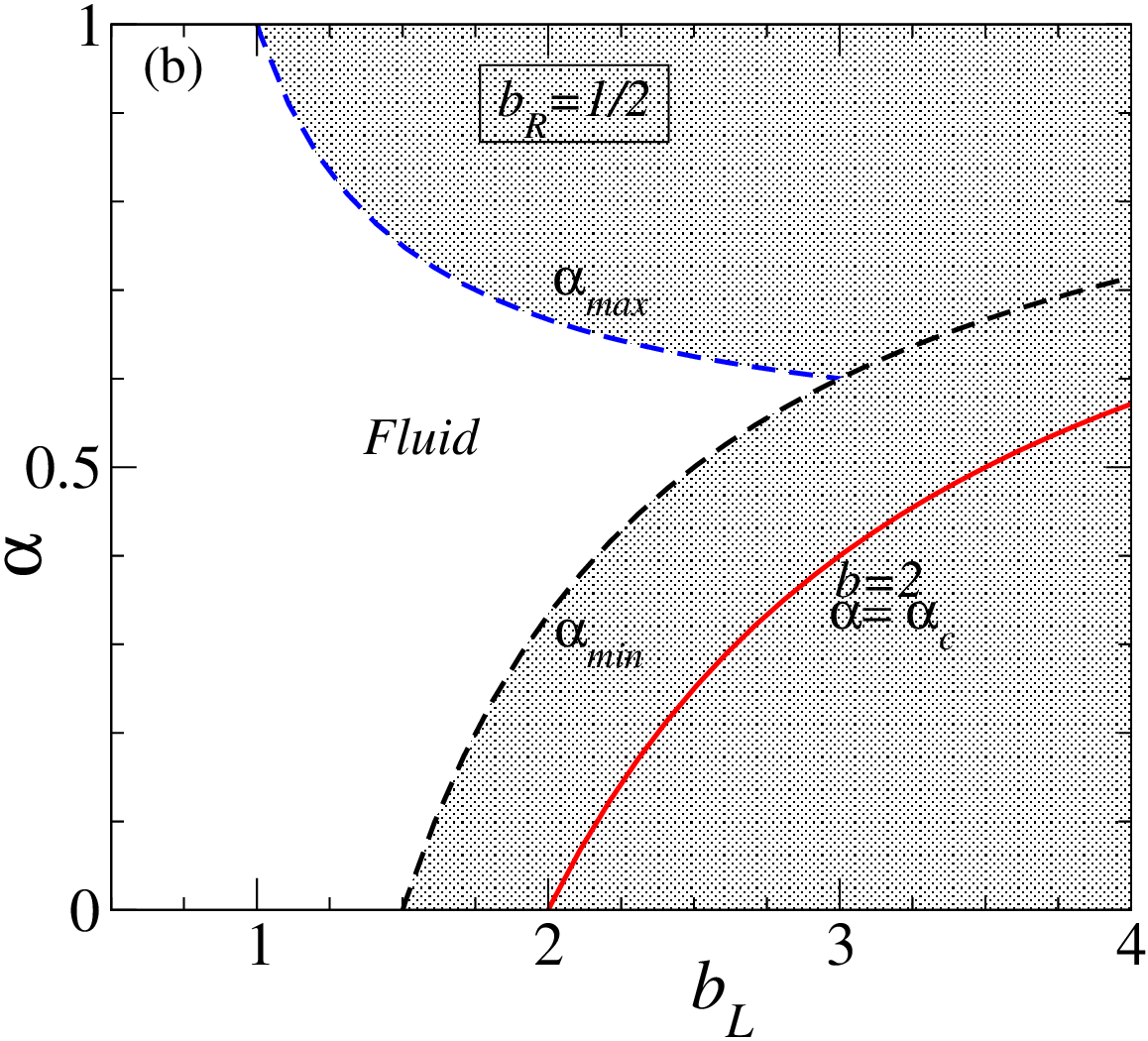}
 \caption{Condensation transition   for  AZRP dynamics  given by Eqs.  (\ref{eq:arl1})-(\ref{eq:arl3}).  For  any given   $b_R,b_L$ the  steady state 
 has a factorized form   when  $ \alpha \in (\alpha_{min},\alpha_{max}).$   Plots  of  $\alpha_{min}$ and $\alpha_{max}$  as function  of $b_L$ ($>b_R$) 
 are shown  here  for  (a) $b_R=  \frac32$ and (b) $b_R=1/2;$  we do not have  exact steady  state solution  in  the shaded  regions where  $\alpha>\alpha_{max}$ 
 or   $\alpha<\alpha_{min}.$    Condensation transition  occurs for large  densities  when   $b = \alpha b_L  + (1-\alpha) b_R$  
 is   larger than $2$. In (a),  this transition line  $b=2,$    which separates the   fluid phase  from the condensate one,  lies in the  region where   
 we have exact  (factorized) steady state.
 }
 \label{fig:1}
\end{figure}
in  Fig.  \ref{fig:1}(a) and (b)  we have  plotted  $\alpha_{min}$ and $\alpha_{max}$  in  dashed lines for  $b_R= \frac 3 2$ and 
$\frac 1 2$  respectively.
The regions  for $\alpha> \alpha_{max}$ and  $\alpha< \alpha_{min}$  are shaded to   indicate 
that within  this  formalism \footnote{Let us   remind that   the condition 
that  we  derive here  is {\it not}  a necessary  but a  sufficient condition.} the steady state   does not  have a   
factorized  form in these regions.   In the rest of regions,  we have a 
factorized steady state  given by  Eqs. (\ref{eq:FSS2}) and  (\ref{eq:1bn}) and  a condensation transition   occurs  here   
for  large  densities ($\rho> \frac{1}{b-2}$)   when  $b$ is greater  than $2,$  which corresponds to $\alpha> \alpha_c$
where 
\be
\alpha_c\,=\,\frac{b_L-2}{b_L-b_R}.
\label{eq:alpha_c}
\ee
In    Fig. \ref{fig:1} we have also shown   $\alpha=\alpha_c$  as a  solid line, marked as $b=2$ and correspondingly $\alpha=\alpha_c$.  
In the left  panel ($b_R=\frac32$)   this line lies  in  the exactly 
solvable regime separating  the  fluid phase from the condensate one.   For $b_R=1/2,$  we could not conclude  if there  is a condensation transition
as   the exact steady state  measure  in the  neighborhood of   $\alpha=\alpha_c$ line is not known.  In fact,  with  some simple  algebra one can show that
for any $1<b_R<2$    the  transition line lies  in the exactly solvable regime, which  is not the case  when $0<b_R\le 1.$ 

As an explicit example, let us  consider $b_R= \frac32, b_L= \frac94;$  in this case  clearly  
$\alpha$ can vary freely in the range $(0,1)$, which  can be seen   from Fig. \ref{fig:1} (a).
The  rate   functions, from  Eq. (\ref{eq:arl1})-(\ref{eq:arl3}), are  now  $ u_R(n) = \alpha \tilde u_R(n),   u_L(n) = (1-\alpha) \tilde u_L(n)$ with 
\bea  \hspace*{-1.5 cm}
\tilde u_R(n) =  \left\{ \ba{cc}    \frac{(13-3\alpha)(2-\alpha)}{2(3-\alpha) }  & n=1 \\     
\frac{(4 n-3\alpha+9)(4n+ 3\alpha-7)} {16 n(n-1)}&n>1 \ea  \right.~~~
\tilde u_L(n) =  \left\{ \ba{cc}   \frac{(13-3\alpha)(3-2 \alpha)}{4(3-\alpha) }  & n=1 \\  
\frac{(4 n-3\alpha+9)(4 n+3\alpha-4)} {16 n(n-1)}&n>1 \ea  \right. \nonumber \eea
It is easy to check that these functions result in the FSS  given by Eq. (\ref{eq:FSS2}) along with 
(\ref{eq:1bn}) where $b = \alpha b_R + (1-\alpha) b_L $.
For $\alpha=1,$   we have  $b= b_R = \frac32$  and  the system remains in the fluid phase for all densities whereas for 
$\alpha =0,$     condensation occurs as  $ b= b_L= 9/4.$   Interestingly for any  arbitrary  $0<\alpha <1$,  
$b= \frac34(3-\alpha)$  and  a  condensation transition  takes place  when  $\alpha$  is decreased 
 below  $\alpha_c= \frac13$ (from Eq.  (\ref{eq:alpha_c})).  For  any $\alpha>\alpha_c,$   the system sets in the 
 condensate  phase only when the density of the system  is  increased  above $\rho_c = \frac4{1-3\alpha}.$   

\subsection{\textbf{Current reversal}}
Another interesting thing that happens in AZRP is the current reversal, where   the direction of current  depends  on 
the particle density of the system. When AZRP with   hop rates $u_{R,L}(n)$  has a factorized steady state given 
by Eq. (\ref{eq:FSS2})  with  $v(n)= u_R(n)+ u_L(n),$ 
the steady state current in the  system can be written  as 
\be
J=  \frac{1}{F(z)} \sum_{n=1}^\infty   [u_R(n)- u_L(n)]  f(n) z^n  =  \la  u_R(n)\ra-\la  u_L(n)\ra \label{eq:J}
\ee
where  $F(z) =\sum_{n=0}^{\infty} z^n f(n).$  As we have discussed,  a sufficient condition  required  for  having a  
factorized steady state in AZRP is  that  $u_{R,L}(n)$    must have  a  form given  by Eq. (\ref{eq:AZRP_generic_form_rates}), with 
some   $0\le \delta\le 1$ and $0\le \gamma\le \delta/ v(n)|_{max}.$ Then $u_R(n)- u_L(n) =   v(n) [2\delta-1 -2 \gamma v(n-1)]$ and thus 
\bea
J&=& (2\delta-1)   \la v(n) \ra  - 2 \gamma  \la  v(n) v(n-1)\ra  \cr &=& (2\delta-1) z  - 2 \gamma z^2. \label {eq:J1}
\eea 
In the  last  step we used  $v(n)= \frac{f(n-1)}{f(n)}$ to calculate $ \la v(n) \ra     =\frac{1}{F(z)} \sum_{n=1}^\infty  v(n) f(n) z^n  =  z$ and 
similarly,  $ \la  v(n) v(n-1)\ra = z^2.$

In a  simple  ZRP with   hop rates $u_R(n) = \alpha  v(n)$ and $u_L(n) = (1-\alpha)  v(n),$  which corresponds to  the choice 
$\delta = \alpha, \gamma=0,$  Eq. (\ref{eq:J1}) leads to $J= (2\alpha-1) z.$ Thus, in ZRP, the  direction of current  $J$  {\it can not} 
be changed  by changing the density  $\rho$ (or equivalently the fugacity $z$); the direction  is   fixed only  by  $\alpha,$ i.e.,  $J$ is positive 
(or negative)  when $\alpha >\frac12$ ($\alpha <\frac12$). The change of  density    can only  increase or decrease  the magnitude of  current, 
it  can not  change the direction of the flow.  But surprisingly  density dependent current reversal is possible in AZRP: 
for a fixed $u_{R,L}(n)$  the direction of  the current  may get reversed when the density of the system is changed.
It is clear from Eq. (\ref{eq:J})  that  such a reversal is not possible when $u_R(n) - u_L(n))$  has the {\it same sign}   for  all  $n>0.$ 
In the following, we  illustrate  with a simple  example  that  direction of   current  can be tuned by the  density,    when  $u_R(n)> u_L(n)$  for all $n$  
except   $n=1$  where   $u_R(n)<u_L(n).$ To this  end, we consider    AZRP with  rate functions  
 \bea
u_R(n) = \left\{ \ba{cc}   \delta  & n=1 \\     \alpha &n>1 \ea  \right.; ~~~~
u_L(n) = \left\{ \ba{cc}   1-\delta  & n=1 \\   1-\alpha&n>1 \ea  \right., \eea
which   follow  Eq. (\ref{eq:AZRP_generic_form_rates}) with  $\alpha = \delta -\gamma$    varying in the range $(0,1)$ and 
 $v(n) =1 ~~\forall~ n>0$   ( and $v(0)=0$).   In this model  isolated particles  
hop   with a different rate than   the rest.   We also  consider $\alpha>\frac12$ and  $\delta<\frac12$   so that  isolated  particles  hop  preferentially  
in a different  direction (here  towards  left) compared to particles from sites having two  or  more particles  which  preferentially move  towards right.
In this case, the flow   direction of current can  depend on  the density  of the system.  For very large density there are only  
few sites   which  contain isolated particles  and the current is  expected   to be  positive (towards right) whereas  for 
very low density most particles are isolated    and  one  expects a   negative current.   Let us see if the direction of the current 
can be reversed  when the density $\rho$ of the  system  falls below a critical threshold  $\rho^*.$ 

Since,  $v(n) =1 ~~\forall~~ n>0$,  this dynamics results in a FSS with $f(n)=1 ~~ \forall~ n \geqslant 0.$ Correspondingly $F(z)=\frac{1}{1-z}$ and 
$\rho= z F'(z)/F(z) = \frac{z}{1-z},$ which in turn implies $z=\frac{\rho}{1+ \rho}.$   Thus  the current,  from Eq. (\ref{eq:J1}), is 
\be
J = \frac{\rho}{(1+\rho)^2} \left[  2\delta -1  +  \rho (2\alpha -1) \right].
\label{eq:cur_AZRP_pq}
\ee
Since $\alpha>\frac12,$ and $\delta< \frac12,$    the current  $J$    flows   in the negative direction  if     density $\rho$  falls below   $\rho^* =   \frac{1- 2\delta}{2\alpha-1}.$

In fact,   it is clear from (\ref{eq:J1})  that  density dependent  current reversal is a   generic feature  of AZRP.  
For generic AZRP with rate functions represented by  (\ref{eq:AZRP_generic_form_rates}),   current reversal  is 
expected  at  fugacity $z^*= \frac{2 \delta -1}{2\gamma}$. But  the crucial point, one must  keep in mind, is   $z^*$ must lie 
 in the range $0 < z^* < v(\infty)$  so that  $z(\rho^*)=z^*$  would  solve for a    physically realizable  density  $\rho*>0.$

 It is worth mentioning that,  at the point of reversal $(z^* ~\mathrm{or~eqivalently}~ \rho^\star),$ the average current $J$ is zero but 
the  steady  state  of the system   is far different from  the  equilibrium  one  which also is characterized by zero current. For  the model we discussed  here,  one 
obtains equilibrium {\it only} for $\delta=\alpha=\frac{1}{2}$ whereas the point of reversal $\rho^\star \,=\, \frac{1-2\delta}{2\alpha-1}$ 
has a finite value  for any  ($\alpha>1/2, \delta<1/2)$ which correspond to a  non-equilibrium scenario as the detailed balance  condition is violated.

\section{Asymmetric  misanthrope  process (AMAP)}
\label{sec:3}
Misanthrope process (MAP) is  an interacting particle system, where hop rate of particles  depends on  both,  the occupation 
of departure site and the arrival site. In contrast to ZRP, here particles  at the departure site not only interact with  other particles 
there, they also explicitly interact  with  particles at the arrival site.  This model can have  a  factorized steady state   in $1d$
if the hop-rate  satisfies  certain condition; for a periodic  lattice with $L$ sites   $i=1,2,\dots, L,$ each site $i$ containing $n_i$ particles, 
if  particles move  to  their right neighbor with rate 
$u(n_i, n_{i+1}),$ the condition for having a  FSS  reads as \cite{evans_beyond_zrp}, 
\be
\hspace*{-.5 cm}
u(m,n)\,=\,u(m+1,n-1)\frac{u(1,m)u(n,0)}{u(m+1,0)u(1,n-1)}\,+\,u(m,0)\,-\,u(n,0). \label{eq:MAP_condition}
\ee
In this section we   generalize the   misanthrope process  to include   asymmetric rate functions $u_{R,L}(., \ast)$,  where the 
subscripts $R,L$  stands for right, left and the arguments ``.'' and ``$\ast$''  correspond to occupation number of departure and arrival sites respectively.
We ask if  the steady  state of  this asymmetric  misanthrope process (AMAP) can be  factorized, and  if so, what would be the corresponding 
condition on the hop-rates ?

\subsection{\textbf{The model and the criterion for FSS}}
Like AZRP, the present section deals with a one dimensional 
periodic lattice with $L$ sites labeled by $i=1,2,\dots L.$ Each site $i$ contains $n_i (\geqslant 0)$ number of particles as earlier 
but the hop rates in AMAP depend not only on the occupancy of the departure site but also on the arrival site. More precisely, a particle 
from a randomly chosen site $i,$ provided $n_i>0$, can either hop to its right neighbor $(i+1)$ with a rate $u_R(n_i,n_{i+1})$ or 
it can move to its left neighbor $(i-1)$  with  a rate $u_L(n_i,n_{i-1})$.

To study whether AMAP  can have  a FSS, as before, we  start  with a  conjecture that the steady state  has a  factorized form  
$ P(\{n_i\})\sim \prod_{i=1}^{L} f(n_i) \delta(\sum_{i=1}^{L}n_i-N) $  and   look for conditions   on the rate functions that 
satisfy $\frac{d}{dt} P(\{n_i\})=0$ in steady state where $P(\{n_i\}),$ the  probability of  each configuration 
$\left\lbrace n_i \right\rbrace,$ follows  the master equation
\begin{equation*}
\bearr{c}
 \frac{d}{dt} P(\{n_i\})\,=\, \sum_{i=1}^{L} [u_R(n_i,n_{i+1})+u_L(n_i,n_{i-1})] \,\,\dots f(n_{i-1}) f(n_{i}) f(n_{i+1})\dots \\ 
 -\sum_{i=1}^{L}u_R(n_{i-1}+1,n_i-1) \dots f(n_{i-1}+1)f(n_i-1)f(n_{i+1})\dots \\ 
 -\sum_{i=1}^{L} u_L(n_{i+1}+1,n_i-1) \dots f(n_{i-1})f(n_i-1)f(n_{i+1}+1)\dots. \\ 
\enarr
\end{equation*}
Let  us  collect all the terms  from the  right hand side of the above equation that  contain both $n_i$ and $n_{i-1}$ as 
arguments of rate functions, and  write them as  $h(n_{i-1})-h(n_i),$  where function $h(.)$ is   yet to be determined, 
 \bea
 u_R(n_{i-1},n_{i})+u_L(n_{i-1},n_{i}) -u_R(n_{i-1}+1,n_{i}-1)\frac{f(n_{i-1}+1)f(n_{i}-1)}{f(n_{i-1})f(n_{i})} \cr
 -u_L(n_{i}+1,n_{i-1}-1) \frac{f(n_{i-1}-1)f(n_{i}+1)}{f(n_{i-1})f(n_{i})}\,=\,h(n_{i-1})-h(n_i) .
 \label{eq:AMAP_f_h}
 \eea
Clearly,  existence of  a  function $h(.)$ ensures that $\frac{d}{dt} P(\{n_i\})= \sum_i h(n_{i-1})-h(n_i)=0.$   Now  
let us   check for  the boundary conditions, i.e. when either of  $n_i, n_{i-1}$ or both  are  zero. Equation (\ref{eq:AMAP_f_h}) is 
automatically satisfied   when   $n_i= n_{i-1}=0.$  
When $n_i=0, n_{i-1} = m>0,$  we have  
\be h(m)=
u_R(m,0) + u_L(m,0) - u_L(1,m-1) \frac{f(m-1)} {f(m)}  
\ee
Here  we have used the facts   that $u_{R,L}(0,\ast)=0$  (particles can not hop from vacant sites), $f(-1)=0$   as $n_i>0$,  
$f(1)/f(0) =1$ (without loss of generality)  and  $h(0) =0$  as  the function  $h(.)$  in Eq. (\ref{eq:AMAP_f_h})  is defined 
up to an arbitrary additive constant. Similarly,  $n_{i-1}=0, n_{i} = m>0$   results in 
\be 
h(m)=u_R(1,m-1) \frac{f(m-1) } {f(m)}.
\ee
Solving the above two equations  for  $f(m)$ and  $h(m)$, 
we  obtain
\bea
h(m) = u_R(1,m-1) w(m)~;~ f(m) = \frac{f(m-1)}{w(m)} =    f(0) \prod_{k=1}^m \frac1{w(k)} \label{eq:AMAP_FSS}\\
{\rm where }~ w(m)=  \frac{u_R(m,0)+ u_L(m,0)}{u_R(1,m-1)  +u_L(1,m-1) }.\nonumber
\eea
Clearly,   for any given  $u_{R,L}(n,m),$   the steady state of AMAP  is  same as that of a simple ZRP with hop rate
$w(m) =\frac{u_R(m,0)+ u_L(m,0)}{u_R(1,m-1)  +u_L(1,m-1)};$ the   function  $w(m),$  however  satisfies $w(1) =1$  (from above definition).
The ZRP correspondence  is  not surprising, as  we know that  a factorized  steady state (\ref{eq:FSS2})   of any model 
can always be obtained from a  simple ZRP   with hop rate $\frac{f(m-1)}{f(m)}.$  Finally  using $f(m)$  and $h(m)$ in Eq. (\ref{eq:AMAP_f_h}) 
we   get  the following condition on hop rates  that  ensures a FSS    in AMAP, 
 \bea
  \hspace*{-1 cm}
  u_R(m,n)&+& u_L(n,m) 
  = \left[  \frac{u_R(m+1,n-1)}{w(m+1)} - u_R(1,n-1)\right] w(n)  +u_R(m,0)\cr
  &+&\left[\frac{u_L(n+1,m-1)}{w(n+1)} -u_L(1,m-1) \right] w(m)+ u_L(n,0).
 \label{eq:AMAP_FSS_constraint}
 \eea
When  particles  move only to right, i.e.  $u_L(.,\ast) =0$ and $u_R(.,\ast) = u(.,\ast)$ this equation reduces  to  the  condition 
Eq. (\ref{eq:MAP_condition})  required     for   the   usual totally asymmetric misanthrope process   to have an  FSS.   
In summary,  a stochastic process on a $1d$ periodic lattice where particles (without obeying hardcore exclusion) hop 
to right or left with different rate functions  $u_{R,L}(m,n)$  that depend  on the occupation numbers $m$ and $n$   of  departure and arrival site
respectively, has a factorized steady state, as  in Eq. (\ref{eq:FSS2})   if  the  rate functions obey Eq. (\ref{eq:AMAP_FSS_constraint}).

Equation  (\ref{eq:AMAP_FSS_constraint}) is  more complicated than  that    the corresponding condition   (\ref{eq:AZRP_constraint})  for AZRP. 
For AMAP   with any  given rate function  $u_{R,L}(m,n)$   one can easily check  if they   obey  Eq. (\ref{eq:AMAP_FSS_constraint}), but 
obtaining  a   generic  form  of  hop rates   that  satisfy this condition  is   rather  difficult.
In the following     we consider   consider a  few special cases.
 A very special class, is the   equilibrium   AMAP. If rate functions  are   related  as follows 
\be
u_L(m,n)\,=\,u_R(n+1,m-1) \,\frac{w(m)}{w(n+1)},
\label{eq:AMAP_gen}
\ee
they   surely satisfy  (\ref{eq:AMAP_FSS_constraint})   required for   having a  FSS, at the  same time 
they also  obey  the  condition  of  detailed balance. Equation (\ref{eq:AMAP_gen})  
clearly describes a class  of generic equilibrium AMAP models in the sense that 
$u_R(n+1,m-1)$  can  still   be chosen freely.    Another  class  of   AMAP  models  that has factorized steady state is 
\be
u_R (m,n) = \delta  u(m,n) +  \gamma u(m,0) u(1,n); \, u_L (m,n) =  \gamma u(m,0) u(1,n).\label{eq:AMAP_genII}
\ee
These rates, when  used in  Eq. (\ref{eq:AMAP_FSS_constraint})  result  in  Eq.  (\ref{eq:MAP_condition}), which is the condition 
required for   an ordinary misanthrope process  with hop rate $u(m,n)$  to have a   FSS.  Thus, Eq. (\ref{eq:AMAP_genII}) describes a
family of  models, parametrized by  two positive  constants $\delta,\gamma$ and   a positive-valued  function   $u(m,n)$   with $u(0,n)=0.$
In  this   case detailed balance is not  satisfied and this class  of models lead to   a unique non equilibrium  steady  state 
having a   factorized from as in Eq. (\ref{eq:FSS2}) with  weight function, 
\be
f(m) = \prod_{k=1}^m \frac{u(k,0)}{u(1,k-1)}.
\ee

In section  \ref{sec:AMAP_cond} we   discuss a  specific  model   of  AMAP  where hop rates 
follow  Eq. (\ref{eq:AMAP_gen}).  In the following section, we   consider a model which   neither satisfies 
Eq. (\ref{eq:AMAP_gen}) nor  Eq. (\ref{eq:AMAP_genII})   but still leads to a    factorized steady state and 
exhibit  density  dependent current  reversal.

\subsection{\textbf{Current reversal in AMAP}}
Like AZRP, it is possible to reverse the direction of the average current $J $ in AMAP, only by tuning the number density $\rho.$ 
Let us consider  the following   rate functions, 
\be
\hspace*{-1cm} u_R(m,n)=\left\{ \ba{ccc} p & n=0& \cr p_1 & n>0,& m=1 \cr p_2 & n>0,& m>1\ea \right. ~;~ 
u_L(m,n)=\left\{ \ba{ccc} q & n=0 &\cr q_1 & n>0, &m=1 \cr q_2 & n>0,& m>1\ea \right.\label{eq:AMAP_cr_rates}
 \ee

It is easy to check that the rates (\ref{eq:AMAP_cr_rates}) satisfy the constraint (\ref{eq:AMAP_FSS_constraint}) only if 
\be q_2 \,= \,p_2 - q + q_1 + \frac{((p + q) q_1)}{(p_1 + q_1)} - \frac{(p (p_1 + q_1))}{(p + q)}  \label{eq:q2} \ee  
With  this choice of  $q_2$ we have a factorized steady state  given by Eq.  (\ref{eq:FSS2})    where  
\be
f(n)\,=\left\{ 
\bearr{cc}1 &  n=0,1 \\
\alpha^{n-1}  & n \geqslant 2\, 
\enarr \right.  ;~  \alpha =\frac{p_1+q_1}{p+q}.
\label{eq:FSS3}
\ee
It is interesting to note that the  steady state weight does not depend on   $p_2;$  any value of $p_2$   generates  the  same   
steady state as  long as $q_2$  defined  in Eq. (\ref{eq:q2})  is  positive.   One must also note that  though the rates in this model obey 
the generic constraint (\ref{eq:AMAP_FSS_constraint}), they do not satisfy detailed 
balance condition and are not in the form of Eq. (\ref{eq:AMAP_gen}), also do not fall in the special class of rates 
given by (\ref{eq:AMAP_genII}).

\begin{figure}[h]
\vspace*{.5 cm}
 \centering
\includegraphics[height=5 cm]{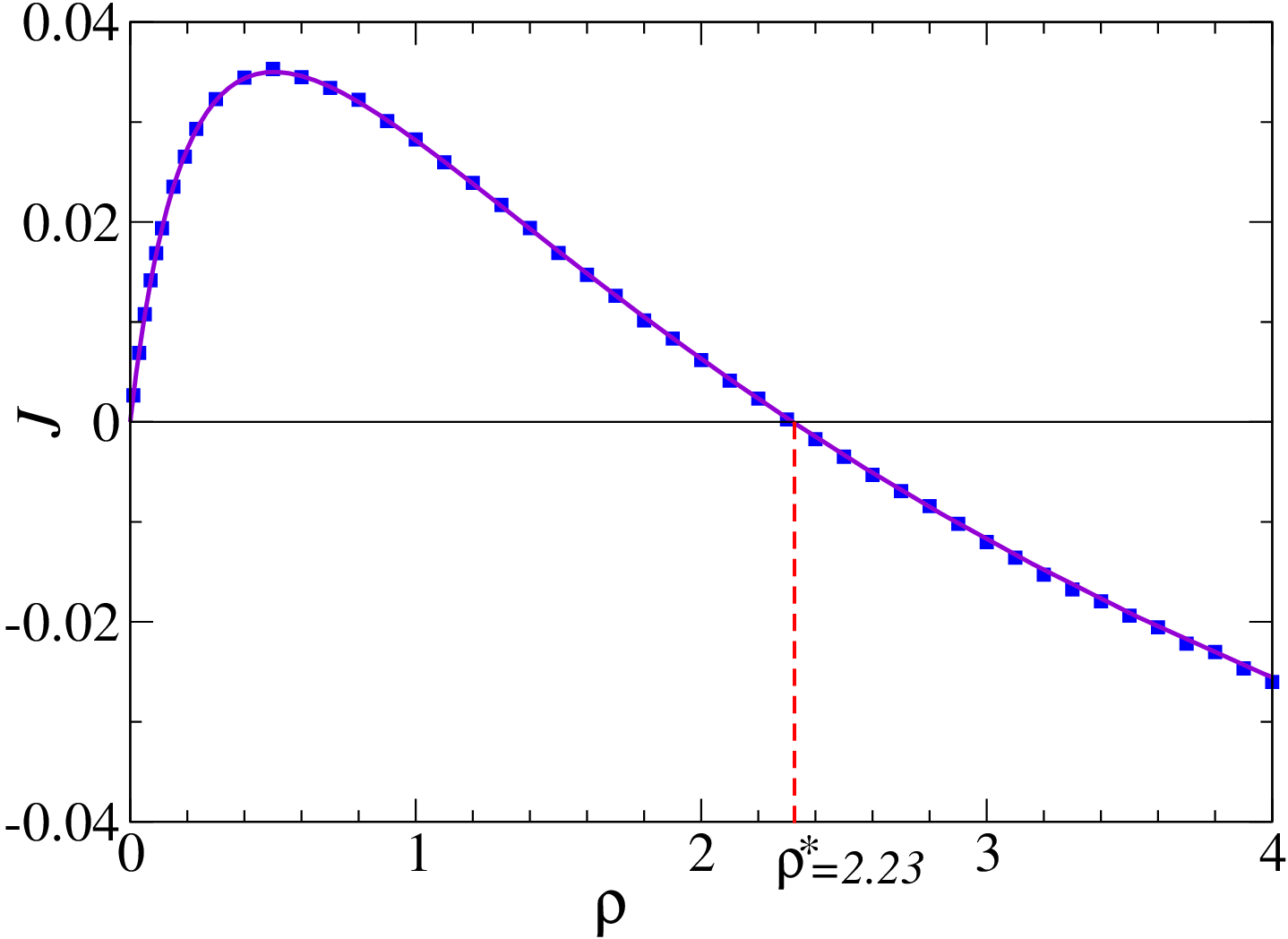} 
 \caption{Current reversal    in AMAP. Current $J$   as  a function of  density $\rho,$   measured  from   Monte Carlo simulation  (symbols)  
 of  AMAP  dynamics   (\ref{eq:AMAP_cr_rates}) with   $(p=\frac12, q = \frac14, p_1 = \frac12, q_1 = \frac34, p_2=53/60, q_2=1)$ 
 on a  system of  size $L,$  is compared with  exact results (lines)  given by Eq. (\ref{eq:AMAP_curr}).  As expected, current reversal occurs at density $\rho^*=2.32.$}
 \label{fig:2}
\end{figure}

In the   grand  canonical  ensemble, the partition function  is   $Z_L= F(z)^L$ with  
$F(z)=    \sum_{n=0}^\infty  f(n) z^n = \frac{1+ (1-\alpha)z}{1- \alpha z}$, where the fugacity  $z$  
lies in the range  $(0,  1/\alpha),$ as the  radius of convergence  of $F(z)$ is $z_c= 1/\alpha.$ 
The  density of the system  is    now 
\bea
  \rho(z) &=& z\frac{F'(z)}{F(z)}=\frac{z}{(1-\alpha z) (1+(1-\alpha) z)} \label{eq:rhopq}\\
{\rm or}~ z &=&  \frac{1+ \rho(2\alpha-1) - \sqrt{ (1-\rho)^2 + 4 \alpha \rho} }{ 2 \rho\alpha (\alpha-1) }. 
\eea
The current  in this system   can be written as 
\bea
J= \frac{1}{F(z)^2} \sum_{m=1}^{\infty}\sum_{n=0}^{\infty}\left[ u_R(m,n)-u_L(m,n)\right]z^{m+n}f(m)f(n)\cr
=\left[ (p-q)\,+\,(p_1-q_1)z\,+\,(p_2-q_2)(F(z)-z-1) \right] \frac{F(z)-1}{F(z)^2} 
\label{eq:AMAP_curr}
\eea
If  $J$  needs to reverse its direction  at some  density  $\rho^*,$   the corresponding   fugacity  $z=z^{\star}$   must be such that 
$J|_{z=z^*}=0;$  using Eq. (\ref{eq:AMAP_curr}) this leads to 
\be
z^*=\frac{1}{\alpha-1} \left[   1- \sqrt{\frac{p(p_1-p) + q(q_1-q)}{p_1q_1-pq} } \right]
\ee
The above value of $z^*$   will correspond to a  feasible density only if $0<z^*<1/\alpha$; and then, one can obtain the corresponding 
density $\rho^* = \rho(z^*)$ using Eq. (\ref{eq:rhopq}).

Now let us consider some specific  cases,  say $\alpha = \frac53.$ This  may be obtained from, say,   $(p=\frac12, q = \frac14, p_1 = \frac12, q_1 = \frac34)$ with  
$q_2= p_2 +\frac7{60}$  (from  Eq. (\ref{eq:q2})).   In this case $z_c = \frac1\alpha = \frac35$ and     the   fugacity  at the  reversal point   $z^*=\frac{3}{4}(2-\sqrt{2}) <z_c.$
So, for this   choice of rates,   current changes its direction when density  of the system   crosses a threshold  value $\rho^* =\rho(z^*)= \frac37(4+\sqrt2)\approx 2.32.$ In 
Fig. \ref{fig:2}, we have shown  a plot of  the average current  as a function of density;  for  very low density  current flows towards  
right and increases as $\rho$ is increased. Beyond a certain density where $J$ reaches its maximum value, it decreases with $\rho$ and 
finally  starts flowing towards  left as  soon as the density becomes larger than  $\rho^*\approx 2.32.$

Another interesting case is $\alpha =1=p+q.$ In this case  when $q_2= p_2+1-2p_1$,   we have a factorized steady state with a weight function 
$f(n) = 1 ~ \forall ~ n >0.$   Thus,  $F(z) =  \frac{1}{1-z},$  and $z=  \frac{\rho}{1+\rho}.$  Now current in the system, from Eq. (\ref{eq:AMAP_curr}),  
\be 
J= \frac{\rho}{(1+\rho)^2} \left[   2p-1 + (2p_1-1)  \rho \right]
\label{eq:cur_AMAP_pp1p2}
\ee
which changes its direction at $\rho^*= - \frac{2p-1}{2p_1-1}.$  Thus  reversal is possible at density $\rho= \rho^*,$  when  
$p> \frac12, p_1<\frac12$ or  when $p< \frac12, p_1>\frac12.$  The noticeable point here is that the current in 
(\ref{eq:cur_AMAP_pp1p2}) is exactly similar to that of the AZRP current in (\ref{eq:cur_AZRP_pq}) with $p\rightarrow \delta$ and 
$p_1\rightarrow \alpha$, so is the point of reversal $\rho^*;$ but the  the  dynamics  or AMAP  is   very different  from  that of  AZRP. 
The similarity  originates from  the  fact  that the stationary state of  both  models are   factorized  with  identical   weight  function 
$f(n) = 1 ~ \forall ~ n \geq 0.$

\subsection{\textbf {Condensation in AMAP} \label{sec:AMAP_cond}}
In this section, we  turn our attention to AMAP models  which give rise to  condensation transition.   
A typical example of such asymmetric rate functions in AMAP that lead to condensation is the following, where 
we consider rates $u_{R,L}(m,n)$ that fall in the special class of AMAP hop rates represented by Eq. (\ref{eq:AMAP_gen}) 
with $w(m) = \frac{1}{1+b} (1+ \frac{b}{m})$ (for $m \ge 1$),
\be
u_L(m,n)\,=\,u_R(n+1,m-1) \,\frac{1+\frac{b}{m}}{1+\frac{b}{n+1}}.
\label{eq:AMAP_condensation_rates}
\ee
This model would result in a FSS   given by Eq. (\ref{eq:FSS2}) along with the single site steady state weight
\be f(n)\,=\,\frac{n! (b+1)^{n}}{(b+1)_{n}},\ee   where $(c)_n=c(c+1)\dots(c+n-1)$is   the  Pochhammer symbol.
  Now,  we can  calculate  the grand canonical  partition function  $Z= F(z)^L$  where 
$F(z) =  \sum_{n=0}^\infty   \frac{n! (1+b)^n}{(1+b)_n} z^n .$   Thus   $z$  varies in  the range $(0,z_c)$ where  $z_c= (1+b)^{-1}$ is the 
radius of convergence   of  $F(z).$  The density of the  system  is now    $\rho(z)= z\frac{F'(z)}{F(z)};$ the critical density  above which condensation 
takes place is 
\be 
\rho_c= \rho(z_c)=\left\{ \ba{cc} \infty   & b \le2 \cr \frac{1}{b-2} & b>2. \ea  \right.\ee 
Thus, for AMAP  with  dynamics 
(\ref{eq:AMAP_condensation_rates}), the system under consideration can  macroscopically distribute  any number of 
particles if $b \leqslant 2$. However, for $b>2$, the maximum allowed density is $\rho_c\,=\,\frac{1}{b-2}$  and  if 
 $\rho$  is  larger than $\rho_c,$  a macroscopic  number,  $(\rho-\rho_c)L$,  of particles gather on some  particular 
site resulting in  the formation of a single site condensate. So, like current reversal, condensation transition is also 
a common feature of both AZRP and AMAP.

\section{\textbf{Asymmetric  finite range process  process (AFRP)}}
\label{sec:4}
Factorized steady state is a very special type of stationary measure but it is not a generic  feature  of the systems out of 
equilibrium. Stochastic processes like ZRP, AZRP, MAP, AMAP constitute a specific class of non-equilibrium processes 
that   enjoy the simplicity of  FSS. But one can also have pair factorized steady state (PFSS) \cite{evans_hanney_majumdar}  
and cluster factorized steady state (CFSS) \cite{FRP} for  generic models  where    particle  interaction   extends   
beyond departure and arrival sites.  Such   finite  range processes   (FRP) introduce spatial  correlations among occupation  
at different  sites  leading to   {\it extended} condensates. Shape and  size of the condensates spreading 
over a finite region in the space  has been extensively   studied in these systems \cite{shape-condensate}.   
In this section, we would like to focus on  asymmetric   FRP in $1d$  where the rate  functions depend
on   occupation of  $K$-nearest neighbors both to right and left of the departure site  but 
the functional form of the hop rates  now  depend on the direction (left or right) of hopping. 
We would like to   find out  specific and sufficient conditions   that must be obeyed by an asymmetric finite range process 
(AFRP) to achieve  a  cluster factorized steady state (CFSS).

\subsection{\textbf{The Model and criterion for CFSS}}
Consider a one dimensional periodic lattice with $L$ sites labeled by $i=1,2,\dots,L$. Each site $i$ contains an integer 
number of particles $n_i(\geqslant 0)$. A particle from a randomly chosen site $i$ (with $n_i>0$) can hop either to its 
nearest right neighbor  $(i+1)$ with rate $u_R(n_{i-K},n_{i-K+1},\dots,n_i,n_{i+1}\dots,n_{i+K-1},n_{i+K})$ or it can hop 
to left nearest neighbor $(i-1)$ at a rate $u_L(n_{i-K}\dots,n_{i-1},n_{i}\dots n_{i+K})$. So both 
the right and left rate functions depend on $(2K+1)$ terms, namely the departure site and its $K$ nearest neighbors  in both 
right and left directions. The  $(2K+1)$  arguments  of  $u_{R,L}(\dots)$ are  spatially  ordered, i.e. 
$1^{st}$  to         $(2K+1)^{th}$   arguments   correspond to  occupancy  of   site $i-K$  to   $i+K$ respectively.
Thus,  $(K+1)^{th}$ argument  corresponds to the occupancy of the departure site $i,$   and   $(K+2)^{th}$  and  $K^{th}$   arguments
are  the  occupancy of the arrival site  for right  and left moves respectively. We  assume  that  a 
cluster factorized steady state is  possible for  AFRP, as given below, and derive consistently 
the constraint required  on the rate functions to obtain such a state.    A    cluster factorized steady state   is represented  by 
 \be
  P(\{n_i\})\sim \prod_{i=1}^{L} g(n_i,n_{i+1},\dots,n_{i+K}) \delta(\sum_{i=1}^{L}n_i-N), 
   \label{eq:AFRP_CFSS}
 \ee
where we call $g(.)$ the cluster weight function that depends on $(K+1)$ variables. 
In the steady state,   with  suitable rearrangement of terms,    the master  equation  of AFRP  can be   written  
as a sum of $L$   terms, each one being a  unique function  $F(.)$  of  $(2K+3)$ arguments $(n_{i-K-1},\dots,n_{i-1},n_i,n_{i+1},\dots,n_{i+K+1}).$
So, in the steady state, 
\be
 \frac{d}{dt} P(\{n_i\}) = \sum_{i=1}^{L} F(n_{i-K-1},\dots,n_{i-1},n_i,n_{i+1},\dots,n_{i+K+1})\,=\,0.
\label{eq:AFRP_ME}
\ee
A sufficient condition   that  satisfy the above  equation (\ref{eq:AFRP_ME}) is when each of the $L$  terms in the right hand side 
individually   vanish, i.e.   $F(n_{i-K-1},\dots n_i\dots,n_{i+K+1})\,=\,0$ for every $i$ $(i=1,2,\dots,L).$  Clearly  this condition 
is   too restrictive and  it is {\it not}  a  necessary condition for  having  CFSS.  We  restrict ourselves to this simple case 
which effectively  leads to, 
\bea 
\hspace*{-1cm} u_R(n_{i-K}&,&\dots,n_i,n_{i+1}\dots,n_{i+K}) + u_L(n_{i-K},\dots,n_{i-1},n_{i}\dots,n_{i+K}) \cr
\hspace*{-1cm}&=&u_R(n_{i-K-1}\dots,n_{i-1}+1,n_{i}-1\dots n_{i+K-1}) \prod_{j=i-K-1}^{i}\frac{g(\tilde{n}_{j},\tilde{n}_{j+1},\dots,\tilde{n}_{j+K})}{g(n_j,n_{j+1},\dots,n_{j+K})} \cr \nonumber
\hspace*{-1cm} &+& u_L(n_{i-K+1}\dots,n_{i}-1,n_{i+1}+1 \dots n_{i+K+1}) \prod_{j=i-K}^{i+1}\frac{g(\hat{n}_{j},\hat{n}_{j+1},\dots,\hat{n}_{j+K})}{g(n_j,n_{j+1},\dots,n_{j+K})} 
\label{eq:AFRP_CFSS_constraint}
\eea
Here $\tilde{n}_{j}=n_j+\delta_{j,i-1}-\delta_{j,i}$ and $\hat{n}_{j}=n_j-\delta_{j,i}+\delta_{j,i+1}$.  This 
constraint (\ref{eq:AFRP_CFSS_constraint}) on the rate functions can be   satisfied   by a family of  hop rates, 
parametrized   by  $\delta>0$ and $\gamma >0,$
\bea
\hspace*{-2cm} u_R(n_{i-K},\dots,n_i,n_{i+1}\dots,n_{i+K})\,=  \delta \frac{g(n_{i-K},n_{i-K+1},\dots,n_i-1)}{g(n_{i-K},n_{i-K+1},\dots,n_i)} \cr
\times \prod_{j=i-K+1}^{i}g(\hat{n}_{j},\hat{n}_{j+1},\dots,\hat{n}_{j+K}) 
  + \gamma \prod_{j=i-K}^{i}\frac{g(\bar{n}_{j},\bar{n}_{j+1},\dots,\bar{n}_{j+K})}{g(n_j,n_{j+1},\dots,n_{j+K})} \cr \nonumber
\hspace*{-2 cm} u_L(n_{i-K},\dots,n_{i-1},n_{i}\dots,n_{i+K})\,= 
 \delta\prod_{j=i-K}^{i-1}g(\tilde{n}_{j},\tilde{n}_{j+1},\dots,\tilde{n}_{j+K}) \frac{g(n_i-1,n_{i+1}\dots n_{i+K})}
{g(n_i,n_{i+1}\dots n_{i+K})} 
\label{eq:AFRP_CFSS_hop_rates}
\eea
where the newly introduced $\bar{n}_j=n_j-\delta_{j,i}$ and $\delta, \gamma$  are constant parameters.

Let us consider   the simplest case   of AFRP, where   particle interaction extends over a   range  $K=1.$ 
In   this case, we  expect a  pair factorized steady state $P(\left\lbrace n_i \right\rbrace)\sim \prod_{i}g(n_i,n_{i+1})  \delta( \sum_{i=1}^L  n_i  -N)$ 
when hop  rates are, 
\bea
\hspace*{-1 cm} u_R(k,m,n)=\frac{g(k,m-1)}{g(k,m)}\left[  \delta g(m-1,n+1)\,
+\,\gamma \frac{g(m-1,n)}{g(m,n)}\right]\cr \hspace*{-.5 cm}
u_L(k,m,n)=\delta  g(k+1,m-1)\frac{g(m-1,n)}{g(m,n)}.
\label{eq:AFRP_PFSS_hoprates}
\eea
Note that  for $\gamma=0,$  the hop rates satisfy detailed balance condition,  and  for 
$\gamma=1, \delta=0,$   we   recover the   usual condition required  for pair factorized state discussed in \cite{evans_hanney_majumdar}.

Also we observe that,  current reversal is not possible   for these particular set of rate functions in Eq. (\ref{eq:AFRP_PFSS_hoprates}) 
which result in pair factorized steady states. This is because,       the    current in these  models turns out to be  $J=\gamma z,$ which  
is just proportional to the fugacity $z$ and  since   density $\rho(z)$ is a monotonic function of $z$, it is not possible to reverse the 
direction of the current by changing $z(\ge 0)$ or equivalently  the density $\rho(z)$.
In fact, for $K > 1$ also  the rate functions in Eq. (\ref{eq:AFRP_CFSS_hop_rates}) give the same average current 
$J = \gamma z$, meaning that there is no current reversal by tuning of  the fugacity or density for these  class of models.
However, the possibility of current reversal with a CFSS produced by asymmetric right-left rate functions 
in one dimension is still not ruled out, because,  to satisfy  the  master equation in the steady state, one may find a 
balance condition different from   the one used here;  then  $J$ may not take such a simple form.

Another common feature of AZRP and AMAP is the formation of condensates which, unlike current reversal, can  
also be observed in case of AFRP within the framework of rate functions given by Eq. (\ref{eq:AFRP_CFSS_hop_rates}). We 
illustrate this briefly with a simple example. For $K=1,$ let us choose 
$g(m,n)=\frac{m+n+1}{(m+1)^b}$, where $b$ is a tunable parameter indicating the onset of condensation. 
The corresponding right-left hop rates are
\bea
u_R(k,m,n) = \frac{k+m}{k+m+1}\left[\delta \frac{m+n+1}{m^b} +\gamma(1+\frac{1}{m})^b\frac{m+n}{m+n+1} \right]\cr\cr
u_L(k,m,n) = \delta \frac{k+m+1}{(k+2)^b}~ (1+\frac{1}{m})^b \frac{m+n}{m+n+1}. \cr
\eea
Using the transfer matrix formalism developed in \cite{FRP}, one can calculate the partition function $Q_L(z)$ in the grand canonical 
ensemble, where $z$  is the fugacity associated with a particle in GCE and subsequently one can also obtain the density $\rho(z)$. Now 
if we proceed to calculate the critical density $\rho_c = \lim\limits_{z\to1} \rho(z)$, we find that for $b \le 4,$ $\rho_c$ diverges 
indicating that the system remains in the fluid phase for $b \le 4$ at any density. Whereas, when $b > 4,$ we have a finite value 
of the critical density given by
\bea
\rho_c&=& \frac{\xi_1(b-1) - 2\xi_2(b)+\xi_3(b)}{2 \xi_2(b)+ 2 \zeta(b-1)\sqrt{\xi_2(b)}  }
+\frac{\zeta(b-2)- \zeta(b-1)   }{ \sqrt{\xi_2(b)}+ \zeta(b-1)  }.
\eea
where   $\xi_k(b) = \zeta(b) \zeta(b-k)$  and  $\zeta(b)$ are Riemann zeta functions. So, for $b>4$, if the density of the system is 
greater than the critical density {\it i.e.} $\rho > \rho_c,$ one can observe a macroscopic number of particles $(\rho-\rho_c)L$ 
gathering at a single but arbitrary lattice site forming a single site condensate.

One can also observe spatially extended condensates in AFRP like the one 
discussed in \cite{evans_hanney_majumdar}, only this time with asymmetric rate functions given by
\bea
\hspace*{-2cm} {\small u_R(k,m,n)=\left\{ \ba{cc} e^{U \delta_{m,1}}[e^{-J(n-m+3)} + e^{-2J}\theta(m-n)+e^{2J}(1-\theta(m-n))] & m\le k,n+2 \cr
e^{U \delta_{m,1}}[e^{-J(m-n-3)} + e^{2J}] & m > k,n+2 \cr
e^{U \delta_{m,1}}[e^{-J(n-m+1)} +  \theta(m-n)+ e^{ 2J} (1-\theta(m-n))] &m > k, m\le n+2 \cr
e^{U \delta_{m,1}}[e^{-J(m-n-1)} + 1] & m \le k,m > n+2
\ea
\right. \nonumber
}\eea
and
\bea
\hspace*{-2cm} u_L(k,m,n)=\left\{ \ba{cc} e^{-J(k-m+3)+ U\delta_{m,1}} & m\le k+2, n \cr
e^{-J(m-k-3)+ U\delta_{m,1}} & m > k+2 , n \cr
e^{-J(k-m+1)+ U\delta_{m,1}} & m\le k+2 ,m> n \cr
e^{-J(m-k-1)+ U\delta_{m,1}} &m > k+2 , m \le n.
\ea
\right. \nonumber
\eea
These rate functions lead to a PFSS with $g(m,n)=e^{-J| m-n | + \frac{U}{2}(\delta_{m,0}+\delta_{n,0})}$. Here $J, U$ are the parameters 
that can be tuned to study the possibility of a condensation transition. As discussed in 
\cite{evans_hanney_majumdar}, when $J>J_c,$ if the density $\rho$ of the system is larger than the critical density 
$\rho_c=\frac{1}{e^{2(J-J_c)}-1}$ (where $J_c=U-\mathrm{ln}(e^U-1)$), a macroscopic number of particles condensate over a spatial 
extent $O(L^{1/2})$ where $L$ is the length of the lattice. 

In brief, we have discussed the possibility of formation of both single site and extended condensates in case of AFRP with $K=1.$

\section{Summary}
\label{sec:5}
We have introduced a  class of one dimensional   stochastic  models of  interacting  particles, without hardcore exclusion,  where the particles are   transferred  
asymmetrically to their neighbors:    both   right and left hop rates    depend on the occupation   of the departure site  and  their neighbors,  
but  their  functional forms  are different.   In  usual driven diffusive systems the asymmetric rate  appears from spatial 
inhomogeneity created by an external potential, which does not depend on  the   microscopic   occupation. However
it is not difficult  to imagine,  in fact actually has been shown recently, through simulations \cite{Geometry} and in biological systems 
\cite{BioChannel,Bacteria}, that  geometric irregularity can result in asymmetric  diffusion of particles. It is interesting to ask what 
kind of   rate functions are realistic for a particular geometry and the answer to this question is not  understood   well.  In this 
article  we  focus  on  generic asymmetric  rate functions and derive   sufficient conditions on them for obtaining exact steady state 
measure   for various asymmetric stochastic processes that include asymmetric  zero-range  process (AZRP), asymmetric   misanthrope process 
(AMAP) and   for the most  generic case,  asymmetric finite range process (AFRP). 

Unlike  ZRP, which has a factorized steady state (FSS) for any  hop rate $u(n)$, AZRP with rate functions $u_{R,L}(n)$   
lead to FSS when the rate-functions satisfy   a specific condition  Eq. (\ref{eq:AZRP_constraint}). On the other hand, a  desired    
FSS  as in Eq. (\ref{eq:FSS2}) can {\it always}  be obtained from  a  two parameter family of AZRP having  left and right hop rates 
described by Eq. (\ref{eq:AZRP_generic_form_rates}).  It is  well known \cite{evans_beyond_zrp, FRP} that  misanthrope 
process can not have a  cluster-factorized steady state  and  its  the   steady state   has a factorized form only for certain   hop rates   
$u(m,n)$ which satisfy  Eq. (\ref{eq:MAP_condition}). AMAP  shares the same feature but with a different constraint on the rate functions; 
it   leads to a FSS only when the hop rates  $u_{R,L}(m,n)$  follow  Eq. (\ref{eq:AMAP_FSS_constraint}). 
Both AZRP and AMAP  show condensation transition, similar to  other models having a FSS.   Interestingly in case of AZRP, 
the condensation transition can be  induced or broken by  tuning the relative choice of  $u_{R,L}(n)$  i.e. by changing the factor that 
decides how often a right move occurs with respect to a left move.
The  important role  of  asymmetric  dynamics, both in AZRP and  AMAP,  appears  in the  particle current. Unlike ZRP or MAP  where the 
direction of  current is fixed by the  external bias, here   the direction  can get reversed by changing particle density.  We also extend 
this  idea of asymmetry between right-left hop rates to  obtain  a  cluster-factorized steady state in AFRP. In particular, we describe 
specific examples where  the  rate  functions  depend  on the occupation of   departure site and  its  two  nearest neighbors (right and left), 
but the functional form  for  the right  hop  is   different from   that of the left; in this  case  we  have  obtained  a sufficient condition  required for  
a pair factorized state. Also, these examples include the formation of both localized and extended condensates.   The general condition   
required  for  AFRP to have  CFSS  is much more complicated and   we could not   obtain 
the most  generic  class  of rates  which satisfy  this constraint. However, in this article, we discuss a specific family of  models 
parametrized by  two constants although they do not   show  density dependent current reversal.  

Some interesting open problems  are   AZRP, AMAP and AFRP with open boundaries or quenched disorder
which may   give rise  to  interesting  boundary driven phase transitions.   In this context, we should mention that
site dependent current fluctuations  above some critical current and that being  indicator of condensation transition for open boundary ZRP with right-left rates related through a multiplicative constant has been studied in detail in \cite{current_fluc}. One can also study the possibility of phase separation in exclusion models corresponding  to the AZRP, AFRP dynamics studied here.

\end{document}